\documentclass[prd,aps,preprint,amsmath,nofootinbib,amssymb,eqsecnum,showkeys,tightenlines]{revtex4-1}
\usepackage{slashed}
\usepackage{epsfig,latexsym,cancel,amssymb,amsmath,verbatim,mathrsfs}
\usepackage{color}
\usepackage{graphicx}

\usepackage{caption}
\usepackage{subcaption}
\usepackage[hidelinks]{hyperref}

\definecolor{My_red}{cmyk}{0.00,1.00,1.00,0.20}

\usepackage{float}%
\usepackage{multirow}

\def\L{\left(}
\def\R{\right)}

\def\ld{\lambda}

\graphicspath{{images/}}

\usepackage[compat=1.1.0]{tikz-feynhand}

\newcommand{\figref}[1]{Fig.~\ref{#1}}
\newcommand{\tabref}[1]{Tab.~\ref{#1}}%

\begin{document}

\title{Interplay between Vector-like Lepton and Seesaw Mechanism:\\
Oblique Corrections}
\author{Shuyang Han}
\email[E-mail: ]{d201980123@hust.edu.cn}
\affiliation{School of physics, Huazhong University of Science and Technology, Wuhan 430074, China}

\author{Zhaofeng Kang}
\email[E-mail: ]{zhaofengkang@gmail.com}
\affiliation{School of physics, Huazhong University of Science and Technology, Wuhan 430074, China}


\author{Jiang Zhu}
\email[E-mail: ]{jackpotzhujiang@gmail.com}
\affiliation{Tsung-Dao Lee Institute and  School of Physics and Astronomy, Shanghai Jiao Tong University,
800 Lisuo Road, Shanghai, 200240 China}
\affiliation{Shanghai Key Laboratory for Particle Physics and Cosmology, 
Key Laboratory for Particle Astrophysics and Cosmology (MOE), 
Shanghai Jiao Tong University, Shanghai 200240, China}

\date{\today}

\begin{abstract}

The non-vanishing neutrino mass strongly hints the existence of right-handed neutrinos (RHNs), singlets of the standard model (SM). However, they are highly decoupled from the SM and difficult to probe. In this work, we consider the Majorana RHNs from the type-I seesaw mechanism may well mix with the heavy neutral lepton dwelling in certain vector-like lepton (VLL), thus acquiring a sizable electroweak charge. Such a simple scenario yields many interesting consequences, and the imprint on oblique corrections, well expected from the mass splitting between components of VLL by virtue of VLL-RHN mixing, is our focus here. We analytically calculate the Peskin-Takeuchi parameters $S$, $T$ and $U$ with full details, carefully treating the Majorana loop to obtain the self consistent expressions free of divergence. Then, we constrain on the VLL-RHN system which only gives a sizable $T$ parameter using the PDG-2021 data and CDF-II data, separately, by imposing $T\lesssim{\cal O}(0.1)$. It is found that for the RHN and VLL below the TeV scale, with a properly large mixing, stands in the frontier of the electroweak precision test such as $W$-boson mass.




\end{abstract}

\maketitle

\newpage


\section{Introduction} 

Although there is a lot of evidence for new physics that goes beyond the standard model of particle physics (BSM), the tiny but non-vanishing neutrino mass is undoubtedly the most convincing one. And the most convincing particle physics model to explain its origin is the seesaw mechanism~\cite{seesaw}. However, how to test the minimal seesaw model, type-I seesaw with at least two right handed neutrinos (RHNs), is somewhat awkward~\cite{Cai:2017mow}, as the RHNs and SM usually are highly decoupled at low energies: they are either very heavy or very weakly coupled to the SM. Considering that the fermionic RHNs may be charged under new gauge interactions, such as $B-L $, opens a way to hunt for them. In this case, the RHNs generically have new gauge and Yukawa interactions, which may help to enhance the production of RHNs at colliders~\cite{Kang:2015nga,Kang:2015uoc,Das:2019fee,Liu:2021akf,Das:2022rbl,FileviezPerez:2020cgn,Accomando:2016sge,Batell:2016zod,Cai:2017mow}.

Is there any other way? A simple idea is to well mix the sterile neutrinos with additional active heavy neutral leptons from the vector-like leptons (VLL) so that the sterile RHNs gain sizable couplings with the weak currents, turning to be the weakly interacting neutral leptons. Such a VLL-RHN system, for various motivations, of course  is not novel and we here mention some of them which are closely related to the current study~\cite{Lavoura:1993mz,Gates:1991uu,Kniehl:1992ez,Ma:1992uc,Lavoura:1992np,Ellis:2014dza,Cynolter:2008ea,Cai:2016sjz,Wang:2022dte,Garg:2013rba,deGiorgi:2022xhr,deGiorgi:2024str}.  The introduction of VLLs may not have a direct relationship with the seesaw mechanism, but even if limited to this framework, there is still sufficient motivation to consider VLL. A good case in point is the family dependent $B-L$ models~\cite{Kang:2019vng,Kang:2020gfi} where VLLs are built-in block, playing the role of flavon field to generate full neutrino mixings. In either case, it is of strong interest to investigate the interplay between VLL and the seesaw mechanism.

In this work, we examine the simplest and also the most interesting VLLs, denoted as $(L_L,L_{R})$, which resemble the SM lepton doublet $\ell_i$. The interplay can be encoded in the Yukawa portal with the Higgs doublet $\lambda_n\bar L_L H N_R$, which has the advantage of not affecting the structure of SM charged lepton flavors. Then, a sizable $\lambda_n$ may lead to a significant interplay between the VLL and RHNs, provided that both of their mass terms are not much larger than the weak scale. In the following, we show several potential consequences in the interplay region:  
\begin{itemize}
    \item The neutral and charged components of VLL gain a large mass splitting due to the mixing with RHN. Then, these states may leave imprints in the oblique corrections.  
    \item The strong mixing between RHNs and the neutral VLL components, which carry full EW charges, results in deep involvement of RHN in electroweak interactions, which has a deep implication to the collider probe on the seesaw mechanism.  
    \item The portal coupling may lead to Higgs invisible decay, given that the RHN is considerably lighter than the weak scale.
    \item The additional Majorana fermions, significantly coupled to the electron via the charged current, may generate a sizable amplitude for $0\nu\beta\beta$-decay, which can be tested by nuclear experiment. 
\end{itemize}


The goal of this work is to anatomize the first point, investigating the allowed region for the VLL-RHN system, which is a necessary preliminary study before exploring other consequences. We will make a detailed calculation of the Peskin-Takeuchi oblique parameters $S$, $T$ and $U$ for this system, where the peculiar nature of Majorana fermions calls for careful treatment. Oblique correction from the VLL-Majorana singlet system is not studied for the first time, and the earliest studies can be traced back to the 1990s, when such a system aroused great interest since it leads to a would-be negative values of $S$ and $T$ favored by data~\cite{Lavoura:1993mz,Gates:1991uu,Kniehl:1992ez,Ma:1992uc}. We also found some related works which deal with the oblique correction from the VLL-Dirac/Majorana system~\cite{Lavoura:1992np,Ellis:2014dza,Cynolter:2008ea,Cai:2016sjz,Wang:2022dte,Garg:2013rba,deGiorgi:2022xhr,deGiorgi:2024str}. Although negative oblique correction is not favored by the current global fit, it is timing to revisit that system because we now have the updated EW precision tests and the powerful LHC.  The PDG-2021 data and as well the recent CDF-II data~\cite{Hays:2022qlw} are able to impose substantial constraints on them. The latter hints a significant deviation to the SM prediction of the $W$-boson mass, but differs significantly from other measurements such as the new ATLAS result~\cite{ATLAS:2023fsi,ATLAS:2024erm}.  We may still need to wait to prove or deny this intriguing anomaly, but anyway, it can be accommodated in a wide range of the VLL-RHN system with $T\sim {\cal O}(0.1)$, especially for a weak scale RHN. Maybe, a more conservative view is to take those range as the forefront of precise testing of sensitivity to VLL-RHN.

The work is organized as the following: In Section II we setup the minimal model for VLL extended type-I seesaw mechanism. In Section III we present the $S,T,U$ calculations.  In the subsequent section numerical result is presented. The final section includes the conclusions and discussions.

\section{Vector-like lepton doublet with a seesaw portal}  

Viewing from the  non-vanishing neutrino masses, the BSM extending the SM by the type-I seesaw mechanism ($\rm SM_{\nu,I}$) is one of the most promising one. But RHNs are sterile and difficult to probe. Our goal is to take $\rm SM_{\nu,I}$ as the basic model (the discussions can be directly generalized to the scenarios of Dirac neutrino), further including VLLs with various motivations that can couple to the seesaw sector. As stated in the introduction, it may deeply alter the profile of RHNs. This section is to set the working model and then present the charged and neutral currents necessary for the subsequent calculations of oblique parameters. 

\subsection{The model setup for \texorpdfstring{$\rm SM_{\nu,I}$}{\rm SM{nu,I}} plus VLLs}
 There are many ways to introduce vector-like fermions with proper electroweak charges and make them become VLLs via certain couplings to the SM leptons. We are not aiming at exhausting the full list of VLLs that contain a neutral component and potentially mix with RHNs, and we focus on the simplest case, a pair of VLL and its charge conjugated states
\begin{align}
    L_{L/R}=\begin{pmatrix} L^0_{L/R}\\L^-_{L/R} \end{pmatrix},\quad L_{L/R}^C=\begin{pmatrix} (L^-_{L/R})^C\\(L^0_{L/R})^C \end{pmatrix},
\end{align}
which like the SM lepton doublet. This option obviously allows for the direct couplings between the VLLs and RHNs. In the following, we will first write down the most general effective model without imposing any symmetry and then a UV model based on the flavorful gauge group $(B-L)$ is given.

\subsubsection{The effective model with simplifications}


For simplicity, we here consider only one family of RHN and lepton, and the generalization to the case with multiple families is straightforward. Then, the most general Lagrangian reads
\begin{align}\label{Lag}
-{\cal L}\supset& \left(y_e\bar{\ell}_LHe_R+y_N\bar{\ell}_L\tilde{H}N_R+h.c.\right)+\frac{M_N}{2}\left( \overline{N_R^C }N_R+ \overline{N_R }N_R^C\right)
\\ \nonumber
&+m_L\bar{L}_LL_R
+m^\ell\bar{\ell}_LL_R+\lambda_n\bar{L}_L\tilde{H}N_R+\lambda_n'\overline{L_R^C}H N_R+\lambda_e\bar{L}_L{H}e_R+h.c.,
\end{align}
where $\tilde H=i\sigma_2 H^{*}$, and the representations of various fields under the electroweak gauge group $SU(2)_L\times U(1)_Y$ can be found in Table.~\ref{QN}. Following the convention widely adopt in the neutrino literature~\footnote{This is in contrast to the convention in supersymmetry~\cite{Dreiner:2008tw}, which describes the right-handed Majorana spinor in terms of the four component spinor $N_R\equiv P_R N_D$ with the upper two components of $N_D$ irrelevant, and the charge conjugated $N_R^C\equiv P_R N_D^C=(N_L)^C$ and $N_L^C\equiv P_L N_D^C=(N_R)^C$.},  we define the charge conjugate of the chiral Weyl spinor as $\psi_{L/R}^C\equiv (\psi_{L/R})^C=C\overline{\psi_{L/R}}^T$ and hence  $\overline{\psi_{L/R}^C }=-\psi_{L/R}^TC^{-1}$ with $C$ the charge conjugation operator. There are useful identities about the bilinear terms holding for the anticommuting spinor fields, for instance,   $\overline {\psi_R}\psi_L=\overline {\psi_L^C}\psi_R^C$,  $\overline {\psi_R^C}\chi_R=\overline {\chi_R^C}\psi_R$, and so on, which are useful to deal with the mixed Dirac-Majorana system. The phase of the Majorana field is chosen to make $M_N$ real and positive.

In Eq.~(\ref{Lag}), the terms in the first line belong to the seesaw mechanism, and in particular, in the limit $M_N\to 0$, we obtain the Dirac neutrino scenario.  The second line gives heavy VLL with a Dirac mass term $m_L$ (generically made real and positive by rephasing the $L_R$), and as well as possible ways that VLL could interact with the seesaw mechanism other than the gauge interactions. VLL interacting with the seesaw model via the RHN portal does not significantly alter the lepton flavor structure of the SM. On the contrary, if $L_L$ is indistinguishable from $\ell_i$, then their sizable mixings will give rise to large LFVs, which have been strongly excluded by the current data. To that end, we set $m^\ell\to 0, \lambda_e\to 0$.  Moreover, for the low scale seesaw, mixing between RHN and SM neutrinos are highly suppressed, and therefore we can safely ignore $y_N$~\footnote{In such a simplified limit, it actually reproduces the well-known bino-Higgsino system in the minimal supersymmetric standard models (MSSM), where the bino is the Majorana fermion and Higgsinos the VLL, but their mixing is limited by the EW gauge coupling. However, this may be changed in the next-to MSSM, where the singlino is a Majorana fermion and the term $\ld SH_uH_d$ allows a larger mixing for $\ld\sim {\cal O}(1)$ favored by enhancing Higgs boson mass.}.

In the above simplified setup, we are considering a sector containing VLL and RHNs only, and then the resulting mass mixing matrix of neutral fermions takes the form of
\begin{equation}\label{3-parameter}
    \mathcal{L}_{\nu\; \text{mass}}=-\frac{1}{2}\overline{\Omega^C_R}
    M\Omega_R+h.c.,\quad
M=\begin{pmatrix}
0&m_L& m_D\\
m_L&0&m_D'\\
m_D & m_D'&M_N
\end{pmatrix},
\;
\Omega_R=
\begin{pmatrix}
    (L^0_L)^C\\L^0_R\\N_R
\end{pmatrix}.
\end{equation}
where $m_D=\lambda_n\frac{v}{\sqrt{2}}$ and $m_D'=\lambda_n'\frac{v}{\sqrt{2}}$. 
These two Dirac mass mixing terms $m_D$ and $m_D'$ mix the neutral component of Dirac VLL with the Majorana RHN, splitting the neutral and charged components. In particular, although the $\lambda_n'$ term is allowed in the effective model setup, it is may be forbidden in the UV-completion model where $N_R$ is not a Majorana spinor but a Weyl spinor carrying charge; we will give such an example soon later.

Using Takagi decomposition one can diagonalize the above complex symmetric mass mixing matrix $M$, via the unitary congruence transformation $V$, i.e., $V^T M V = \hat{M}$, and then the mass terms become
\begin{equation}\label{three masses}
    \mathcal{L}_{\nu\; \text{mass}}=-\frac{1}{2}
    \overline{\omega^C_R}\hat M\omega_R+h.c.,
\end{equation}
where $\omega_{R_i}$ are three massive right-handed Weyl states in the mass basis, and they are related to $\Omega_{Ri}$ as the following 
\begin{equation}\label{V}
    \Omega_R^C=
    \begin{pmatrix}
        L^0_L\\(L^0_R)^C\\N_R^C
    \end{pmatrix}=V^*\omega_R^C,\quad 
    \Omega_R=
    \begin{pmatrix}
        (L^0_L)^C\\L^0_R\\N_R
    \end{pmatrix}=V\omega_R ,
\end{equation}
where the neutral mass spectrum are arranged as $\hat{M}={\rm diag}(M_1,M_2,M_3)$ with $M_i$ real and positive, ordered from light to heavy.

To switch to the 4-component notation, we embed $\omega_R$ into  the 4-component Majorana fields as $N\equiv \omega^C_R+\omega_R$, which obviously satisfies the self-conjugate constraint $N=N^C$; moreover, one has $N_R=P_RN$ and $N_L=P_LN\equiv N_R^C$ with $P_{L,R}=(1\mp \gamma_5)/2$. Then, the mass terms in  Eq.~(\ref{three masses}) can be combined into a single term,  $\mathcal{L}_{\nu\; \text{mass}}=-\frac{1}{2}M_i\overline{N_i}N_i$.

The large mixings between VLLs and RHNs may affect the active neutrino phenomenology in the seesaw mechanism, since the mixings bring the two neutral VLLs to the neutrino mass mixing matrix. This leads to the modified seesaw sector containing two more heavy Majorana fermions which form the light-heavy Dirac mass terms $(y^{eff}_N)_{ai}\bar{\ell}_a\tilde{H}N_i$ with $(y^{eff}_N)_{ai}=(y_N)_{a1}V_{3i}+(y_N)_{a2}V_{4i}$, where the flavor indices in $(y_N)_{a1,2}$ refer to $\ell_a$ and $(N_R)_{1,2}$ in a realistic model, and $V_{3,4i}$ come from $(N_R)_{1,2}=V_{3,4i}N_i$. Then, the resulting effective mass mixing matrix for the active neutrinos is
\begin{align}
    (m_\nu^{eff})_{ab}=-v^2(y^{eff}_N)_{ai}M_i^{-1}(y^{eff}_N)_{ib}.
\end{align}
Hence, the type of Majorana and Dirac mixing does not relax the requirement that at least two RHNs are needed to produce two non-vanishing active neutrino masses.

\subsubsection{A UV completion in the local \texorpdfstring{$(B-L)_{ij}$}{(B-L){ij}} model}

The structure of the previous effective model is naturally realized in the flavorful local $B-L$ extension to the SM. It is well-known that the fermions within the SM are subject to anomaly cancellation, while beyond the SM, the appearance of RHNs can be elegantly ascribed to the gauged $(B-L)$. Minimally, we only need two RHNs, corresponding to the local $(B-L)_{ij}$ for the $i,j$-th generations of fermions~\cite{Kang:2019vng,Kang:2020gfi}; for concreteness, we here consider $(B-L)_{13}$ and other options are similar~\footnote{Ref.~\cite{Kang:2019vng,Kang:2020gfi} studied $(B-L)_{23}$ in order to get a light gauge boson $Z_{B-L}$ phobic to electron, and therefore  $Z_{B-L}$ is able to enhance $(g-2)_\mu$, avoiding the exclusions from many low energy experiments.} .

Although successfully explaining the neutrino masses, the minimal $(B-L)_{13}$ models can not fully account for the Pontecorvo-Maki-Nakagawa-Sakata (PMNS) matrix~\cite{Maki:1962mu,Pontecorvo:1967fh},  owing to the flavorful $B-L$. To realize the PMNS mixings, we minimally introduce a pair of vector-like lepton doublet $L_{L,R}$ and as well as a flavon $\mathcal{F_\ell}$, required to develop a non-vanishing VEV via a proper scalar potential; their quantum numbers in the new gauge group can be found in Table.~\ref{QN}. Given the above field content and symmetries, the most general Lagrangian of the leptonic sector reads 
\begin{align}\label{model:B-L}
-\mathcal{L}_L &=Y^e_{22}\bar{\ell}_{2}H e_{R2}+Y^{e}_{ij}\bar\ell_{i}He_{Rj}+Y^{N}_{ij}\bar\ell_{i}\tilde{H}N_{Rj}+Y^e_i \overline{L}_L H e_{Ri} + Y^N_i \overline{L}_L \widetilde{H} N_{R i}
\\
\nonumber
&+\lambda_2^{\ell} \bar{\ell}_2 L_R \mathcal{F}_{\ell}^* +
M^{\ell}_i \bar{\ell}_i L_R + m_L \overline{L}_L L_R+\frac{\lambda^N_{i j}}{2}\Phi \overline{ N^{C}_{R i}}N_{R j} + h.c.,
\end{align}
with the Latin indices $i/j=1,3$ and $a/b=1,2,3$, the mixing coupling $\lambda_n$ in the effective model is replaced with $Y^N_i$ here. $\Phi$ is a Higgs field to spontaneously break $(B-L)_{13}$  and give large Majorana mass to $N_{Ri}$. Note that the term $\lambda_n'\overline{L_R^C}H N_R$ present in the effective model is not present here, due to the  $(B-L)_{13}$ selection rule. 
\begin{table}[htbp]
\begin{center}
\begin{tabular}{|c|c|c|c|c|c|c|c|c|}
\hline\hline
  & $\ell_{e},\ell_{\mu},\ell_{\tau}$ & $e_{R},\mu_{R},\tau_{R}$ &  $N_{e R},N_{\tau R}$ & $H$ & $\Phi$ & $L_{L/R}$  & $L_{L/R}^C$  & $\mathcal{F}_{\ell}$  \\
\hline
SM & (2,-1) & (1,-2) & (1,0) & (2,1)&(1,0) & (2,-1) & (2,1)  & (1,0) \\
\hline
${B-L}$ & -1,0,-1 & -1,0,-1 & -1,-1  & 0 & 2 & -1 & 1 &-1\\
\hline\hline
\end{tabular}
\end{center}
\caption{Field content and quantum numbers in the SM (first) and in the $({B-L})_{13}$ extension (second); the first generation of fermions are neutral under this new gauge group. $L_L$, $L_R$ and ${\cal F}_{\ell}$ are new particles for realizing active neutrino mixings.}
\label{QN}
\end{table}

\subsection{Charged/neutral current couplings}

The main task of this article is to investigate the hints of VLL in the scenario of seesaw extension of SM in the oblique parameters, which are calculated from the EW currents. Let us combine the two chirality $L_L$ and $L_R$ into the single Dirac field $L^T=(L^0,L^-)$, and then in the interacting basis, the charged current is written as 
\begin{align}
\mathcal{L}_{CC}
=&\frac{g}{\sqrt{2}}W^+_{\mu}\left(
\begin{pmatrix}
\bar{\nu}_{a}&\overline{L^0}
\end{pmatrix}
\gamma^{\mu}P_R 
\begin{pmatrix}
0_{aa}&0\\0&1
\end{pmatrix} 
\begin{pmatrix}
e_{a}\\L^-
\end{pmatrix}
+ 
\begin{pmatrix}
\bar{\nu}_{a}&\overline{L^0}
\end{pmatrix}\gamma^{\mu}P_L  
\begin{pmatrix}
\mathbb{I}_{aa}&0\\0&1
\end{pmatrix}
\begin{pmatrix}
e_{a}\\L^-
\end{pmatrix}\right)+h.c.,
\end{align}
where we have introduced the block matrices to label the extended generation structure in the left- and right-handed currents.  Although the SM leptons (with family indices $a=1,2,3$) are irrelevant to our present discussion, they are still included, for our subsequent study elsewhere. Moreover, without loss of generality, we work in the basis that the SM leptonic Yukawa couplings are in the flavor diagonal basis. The neutral current reads
\begin{align}
\mathcal{L}_{NC} \supset&
    \frac{g}{2c}Z_{\mu}
    \begin{pmatrix}
        \bar{\nu}_{a}&\overline{L^0}
    \end{pmatrix}\gamma^{\mu}P_{R}
    \begin{pmatrix}
        0_{aa}&0\\0&1
    \end{pmatrix}
    \begin{pmatrix}
        {\nu}_{a}\\L^0
    \end{pmatrix} +\frac{g}{2c}Z_{\mu}
\begin{pmatrix}
     \bar{\nu}_{a}&\overline{L^0}
\end{pmatrix}\gamma^{\mu}P_{L}
\begin{pmatrix}
        \mathbb{I}_{aa}&0\\0&1
\end{pmatrix}
\begin{pmatrix}
        {\nu}_{a}\\L^0
\end{pmatrix}\notag\\ 
&
+ \frac{g}{2c}(s^2-c^2)Z_{\mu}
\begin{pmatrix}
        \bar{e}_a&\overline{L^-}
\end{pmatrix}\gamma^{\mu}P_{L}
\begin{pmatrix}
        e_a\\L^-
\end{pmatrix}  \notag\\ 
&
+
\frac{g}{2c}Z_{\mu}
\begin{pmatrix}
        \bar{e}_a&\overline{L^-}
\end{pmatrix}\gamma^{\mu}P_{R}
\begin{pmatrix}
        2s^2&0\\0&s^2-c^2
\end{pmatrix}
\begin{pmatrix}
        e_a\\L^-
\end{pmatrix},
\end{align}
where $s^2\equiv \sin^2\theta$ with $\theta$ the weak mixing angle. 

Now, let us rotate the states into the basis defined in the previous subsection, via the unitary transformation Eq.~(\ref{V}). Then, the charged current becomes 
\begin{align}
\mathcal{L}_{CC}
=&\frac{g}{\sqrt{2}}W^+_{\mu}
\begin{pmatrix}
   \bar{\nu}_{a}&\overline{N_1}&\overline{N_2}&\overline{N_3}
\end{pmatrix}
\gamma^{\mu}\left( P_R V_R+P_LV_L \right)
\begin{pmatrix}
\ell_{a}\\e_4
\end{pmatrix}
 +h.c.,
\end{align}
where we have used $N_{Ri}^C=P_LN_i$ to eliminate the notation of charge-conjugated fields. The neutral states are decomposed into first three SM active neutrinos $\nu_a$ plus three heavy $N_i$, and the charged states consist of three SM charged leptons $\ell_a$ plus one heavy $e_4$. In the second term of the first line, we have used $N_{Ri}^C=P_LN_i$. Correspondingly, we introduce the following 6-to-4 flavor mixing matrix $V_{L,R}$
\begin{equation}
V_L=
\begin{pmatrix}
\mathbb{I}_{3\times 3}&0_{3\times1}\\0_{3\times3}&V^{T}
\begin{pmatrix}1\\0\\0\end{pmatrix}
\end{pmatrix},\;
(V_R)_{n4}=
V^*_{2(n-3)}, n=4,5,6,
\end{equation}
with other elements of $V_R$ being zero. We have not considered the light-heavy mixing that leads to the SM neutrino masses yet, so no PMNS mixings appear. The neutral current couplings take the form 
\begin{align}
\mathcal{L}_{NC} =&
\frac{g}{{2}c}Z_{\mu}\left(
\begin{pmatrix}
   \overline{N_1}&\overline{N_2}&\overline{N_2}
\end{pmatrix}\gamma^{\mu}P_{R}W_1
\begin{pmatrix}
   N_1\\N_2\\N_3
\end{pmatrix}+
\begin{pmatrix}
   \overline{N_1}&\overline{N_2}&\overline{N_3}
\end{pmatrix}\gamma^{\mu}P_{L}W_2
\begin{pmatrix}
   N_1\\N_2\\N_3
\end{pmatrix}\right)\notag\\
&+\frac{g}{{2}c}Z_{\mu}\bar{\nu}_{a}\gamma^{\mu}P_L\nu_{a}
+    
\frac{g}{2c}(s^2-c^2)Z_{\mu}
\begin{pmatrix}
     \bar{\ell}_{a}&\bar{e}_4
\end{pmatrix}\gamma^{\mu}P_{L}
\begin{pmatrix}
    \ell_{a}\\e_4
\end{pmatrix}\notag\\
&+
\frac{g}{c}s^2 Z_{\mu}
\begin{pmatrix}
     \bar{\ell}_{a}&\bar{e}_4
\end{pmatrix}\gamma^{\mu}P_{R}
\begin{pmatrix}
     \ell_{a}\\e_4
\end{pmatrix}+
\frac{g}{2c}Z_{\mu}
\begin{pmatrix}
     \bar{\ell}_{a}&\bar{e}_4
\end{pmatrix}\gamma^{\mu}P_{R}W
\begin{pmatrix}
     \ell_{a}\\e_4
\end{pmatrix}.
\end{align}
To write the flavor-changing-neutral-current (FCNC) couplings in a compact form, we have introduced three flavor mixing matrices $W$, $W_1$ and $W_2$ as the following
\begin{align}
W_1=
V^\dagger
\begin{pmatrix}
0&0&0\\0&1&0\\0&0&0
\end{pmatrix}V,\quad
W_2=
V^T
\begin{pmatrix}
1&0&0\\0&0&0\\0&0&0
\end{pmatrix}V^*,\quad W=U_{R}^{\dagger}
\begin{pmatrix}
     0_{3\times 3}&0_{3\times 1}\\0_{1\times 3}&-1
\end{pmatrix}U_{R},
\end{align}
where $U_{R}$ is the $4\times 4$ unitary transformation matrix acting on the right-handed charged leptons, and here we do not consider the heavy-light mixing and thus the only non-vanishing element of $W$ is $W_{44}=-|U_{R}(4,4)|^2=-1$.

Actually, in calculating the obliques parameters $S$, $T$ and $U$, the relevant current couplings involve only the heavy states, and we list them below: 
\begin{align}\label{CC}
\mathcal{L}_{CC}\supset&
\frac{g}{\sqrt{2}}W^{+}_{\mu}
\overline{N_a} \gamma^{\mu}\left( P_R V^*_{2a} +P_L V_{1a} \right) e_4+h.c.,\\
\mathcal{L}_{NC}\supset
&\frac{g}{2c}Z_\mu\overline{N_a}\gamma^{\mu}\left(V_{1a}V^*_{1b}P_L+V^*_{2a}V_{2b}P_R\right){N_b}
+    
\frac{g}{2c}(s^2-c^2)Z_{\mu}
\bar{e}_4 \gamma^{\mu}e_4.\label{NC:1}
\end{align}
However, the above expression does not reflect the fact that $N_a$ are Majorana spinor fields. To implement it, one can transpose the corresponding Lagrangian terms and use the Majorana constraint $N_a^C=N_a$ ~\cite{Gates:1987ay,Dreiner:2008tw,Lavoura:1992np,Lavoura:1993mz,Gates:1991uu,Kniehl:1992ez,Ma:1992uc,Gates:1987ay}, to write the neutral current couplings as (for more details please see Appendix.~\ref{Majorana:interaction})
\begin{align}\label{NC}
\mathcal{L}_{NC}\supset
&\frac{g}{c}Z_\mu\overline{N_a}\gamma^{\mu}\left((g_L)_{ab}P_L+(g_R)_{ab}P_R\right){N_b}
+    
\frac{g}{2c}(s^2-c^2)Z_{\mu}
\bar{e}_4 \gamma^{\mu}e_4, 
\end{align}
with the new couplings encoding the Majorana nature 
\begin{equation}\label{1/2factor}
    (g_L)_{ab}=\frac{V_{1a}V^*_{1b}-V^*_{2b}V_{2a}}{4},\quad
    (g_R)_{ab}=\frac{V^*_{2a}V_{2b}-V_{1b}V^*_{1a}}{4}.
\end{equation}
Note that in Eq.~\eqref{NC:1}, both numerical coefficients for the Majorana fermions and Dirac fermions are $\frac{1}{2}$, denoting for the $T_3$ charge. But in Eq.~\eqref{1/2factor}, the former is multiplied by an additional $\frac{1}{2}$ factor for the sake of satisfying the Majorana condition and becomes $\frac{1}{4}$.

When calculating the $S$ parameter, involving the neutral current couplings only, it is more convenient to work in the interacting basis for the gauge bosons - $W^3$ and $B$:
\begin{equation}
\begin{split}\label{new basis}
 \mathcal{L}_{NC}\supset 
 gW^3_{\mu}\left(
\overline{N_a}\gamma^{\mu}\left((g_L)_{ab}P_L+(g_R)_{ab}P_R\right){N_b}
 -  \frac{1}{2}\bar{e}_4 \gamma^{\mu} e_4\right)
 \\
 -g'B_{\mu} \L
\overline{N_a}\gamma^{\mu}\left((g_L)_{ab}P_L+(g_R)_{ab}P_R\right){N_b}
+   
\frac{1}{2}\bar{e}_4 \gamma^{\mu} e_4\R.
\end{split}
\end{equation}
which is readily obtained from the quantum number of the fields. More details about the currents and their relations in different basis or conventions can be found in Appendix.~\ref{ps.various expressions}.

\section{Calculation of the Peskin-Takeuchi parameters}

The electroweak precise observables (EWPOs) provide a promising way to search for clues to new physics, in particular for the heavier new resonances that can not be abundantly produced at the colliders with distinguishable signatures. In this framework, one expresses the theoretical prediction of an EWPO $\mathcal{O}^{}$ as the SM prediction $\mathcal{O}_{\rm SM}$ plus oblique corrections, which are some linear combinations of the self energies of the EW gauge bosons contributed by new physics and expected to slightly shift $\mathcal{O}_{\rm SM}$. Then, $\mathcal{O}^{}= \mathcal{O}_{\rm SM}\left(1+\text{oblique corrections}\right)$. In the linear approximation, it is sufficient to parameterize the oblique corrections in terms of three Peskin-Takeuchi oblique parameters $S$, $T$ and $U$~\cite{Peskin:1990zt,Peskin:1991sw}. In the following, we will first briefly review them and clear the conventions for their calculations. Then, we present our results specific to the VLL-RHN model.

\subsection{Definition and convention}

The new physics information can be encoded in the oblique corrections to the electroweak propagators, namely the vacuum polarization amplitude between two vector bosons, $\Sigma^{\mu\nu}_{VV'}\equiv \langle J^{\mu}_VJ^{\nu}_{V'}\rangle$; $V$ and $V'$ denote the gauge bosons of the electroweak sector, and they can either be $W^\pm, Z, \gamma$ or $1(2), 3, Y (Q)$, depending on the basis, with details cast in the Appendix.~\ref{ps.various expressions}. The Fourier form $\Sigma^{\mu\nu}_{VV'}(q^2)$ is simply obtained by multiplying $-i$ to the amplitude of the corresponding loop diagram. Only the transverse part is involved, $\Sigma_{VV'}^{T}(q^2)=\frac{1}{3}P_{\mu\nu}\Sigma_{VV'}^{\mu\nu}(q^2)$ with the projector $P_{\mu\nu}=g_{\mu\nu}-q_\mu q_\nu/q^2$. The superscript ``$T$" will be dropped hereafter.  For the low energy experiments with a low $q^2$ such as $q^2= M_Z^2$, it is proper to expand $\Sigma_{VV'}(q^2)$ around $q^2=0$ and take the linear approximation, 
\begin{equation}
    \Sigma_{VV'}(q^2) \simeq \Sigma_{VV'}(q^2=0) + q^2\Sigma'_{VV'}(q^2=0).
\end{equation}
Peskin and Takeuchi found that it is sufficient to parameterize the oblique corrections in terms of three parameters $S$, $T$ and $U$~\cite{Peskin:1990zt,Peskin:1991sw}
\begin{equation}\label{STU parameterization}
\begin{split}
S&\equiv {16\pi}[\Sigma'_{33}(0)-\Sigma_{3Q}'(0)],
\\ 
T&\equiv\frac{4\pi}{M_Z^2s^2c^2}[2\Sigma_{11}(0)-\Sigma_{33}(0)],
\\ 
U&\equiv 16\pi[2\Sigma'_{11}(0)-\Sigma'_{33}(0)],
\end{split}
\end{equation}
where the expanding point $q^2=0$ will be implied. For $S$, it is more convenient to calculate in the basis before EWSB, which takes the form of
\begin{align}\label{S parameterization}
S&=
-16\pi\Sigma'_{3Y},
\end{align} 
where the relations in Appendix.~\ref{ps.various expressions} have been used. 

Usually, the value of the $U$ parameter is very close to zero and not sensitive to new physics. The $T$ parameter is the only one receiving leading order contribution and dominates in the oblique corrections to EWPOs. $T$ parameter is equivalent to the $\rho$ parameter, $T=(\rho-1)/\alpha$, and hence it is sensitive to the violation of custodial symmetry, e.g., due to mass splitting between the isospin multiplets. In the absence of such violations, the $S$ parameter will take over. In convention, the oblique parameters are defined to be zero within the SM. Therefore the SM is just a reference point and these three oblique parameters are indications for BSM, i.e., $O=O|_{NP}-O|_{SM}$ with $O=S, T, U$.

The BSM contribution to the vacuum polarization amplitude $\Sigma^{new}_{VV'}(q^2)$ is shown in \figref{systemic Feynman diagram for W mass contribution}. The calculation is based on the couplings between the neutral or/and charged currents and gauge bosons, which are  generically denoted as $ J \bar gV$ with $\bar{g}=
\tilde g (aP_L+bP_R)$. 
In this convention, we extract out the gauge couplings $\tilde g=g, g', e$ of gauge groups $SU(2)_L$, $U(1)_Y$ and $U(1)_{em}$, respectively, which enables us to write the oblique corrections in the form of Eq.~\eqref{STU parameterization} and Eq.~\eqref{S parameterization}. The concrete expressions of $a$ and $b$ in our model can be read from Eq.~(\ref{CC}), Eq.~(\ref{NC}) and Eq.~(\ref{new basis}). 
\begin{figure}[H]  
\centering    
\begin{minipage}{0.4\linewidth}
    \begin{tikzpicture}[line width=1.0 pt, scale=0.6, >=latex]
    \begin{feynhand}
    \vertex (a) at (-5,0) ;
    \vertex [label=right:$\bar{g}$] (b) at (-2,0) ;
    \vertex [label=left:$\bar{g}'$] (d) at (2,0) ;
    \vertex (e) at (5,0) ;
    \propag  [boson] (a) to [edge label = $V$](b);
    \propag  [boson] (d) to [edge label = $V'$](e);
    \draw (2,0) arc (0:360:2);
\node at (0,2.4) {$\ell$};
\node at (0,-2.4) {$\ell'$};
\end{feynhand}
    \end{tikzpicture}
\end{minipage}
    \caption{Vacuum polarization amplitudes for the electroweak vector bosons $V/V'$; $\ell$ and $\ell'$ denote the BSM fermions like $e_4$ and the heavy neutral leptons.}
    \label{systemic Feynman diagram for W mass contribution}
\end{figure}

We first deal with the leading part in the linear approximation, $\Sigma^{new}_{VV'}(q^2=0)$, which is simply denoted as a scalar function $\Sigma(a,b,m,M)$ with $a,b$ the couplings and $m,M$ masses of the loop particles; the subscripts $VV'$ correspond one-to-one with $(a,b)$, so they can be omitted in this notation. For the case with $V=V'$,  $\Sigma$ can be written as
\begin{align}\label{sigmamassandcoupling}
\Sigma =&(|a|^2+|b|^2) \Sigma_{V+A} + 2\text{Re}(ab^*)\Sigma_{V-A}.
\end{align}
where the subscripts $V \pm A$ denote amplitudes of chiral-hold and chiral-mixing (or $LL/RR$ and $LR/RL$). While for the case with $V\neq V'$ such as $\gamma$-$Z$, $3$-$B$ and  $3$-$Q$, it takes the form of 
\begin{align}
    \Sigma =&(a_{1}a_{2}^*+b_{1}b_{2}^*) \Sigma_{V+A}+(a_{1}b_{2}^*+b_{1}a_{2}^*)
    \Sigma_{V-A},
\end{align}
with ``1" and ``2" denoting for $V$ and $V'$. In this notation, $\Sigma_{V\pm A}$ are just functions of loop masses. For the derivatives part ${\Sigma'}(q^2=0)$, one has similar expressions. 

For the vector couplings to the vector bosons, one has $a=b$, and then it is convenient to introduce $\tilde{\Sigma}=\Sigma_{V+A} +\Sigma_{V-A}$, so we have
\begin{equation}\label{vector-typesigma}
\Sigma=2|a|^2\tilde{\Sigma}=2|b|^2\tilde{\Sigma}.
\end{equation} 
In our model, it is the case for the neutral current couplings of $e_4$. 


Now, we give the concrete expressions for the one-loop functions introduced in the above equations, 
\begin{align}\label{sigmamass}
\Sigma_{V-A}(m,M)=&\frac{1}{8\pi^2}mM \left(\log\frac{mM}{\mu^2}+\frac{(m^2+M^2)}{2(m^2-M^2)}\log\frac{m^2}{M^2}-1\right),\\
\Sigma_{V+A}(m,M)=&-\frac{1}{8\pi^2}\left(\frac{m^4+M^4}{4(m^2-M^2)}\log\frac{m^2}{M^2}+\frac{m^2+M^2}{2}\log\frac{mM}{\mu^2}-\frac{m^2+M^2}{4}\right).
\end{align}
For the derivatives part, the loop functions are given by
\begin{align}\label{sigmaprimemass}
\Sigma'_{V-A}(m,M)=&  -\frac{1}{8\pi^2}m M \left(\frac{m^2+M^2}{2 \left(m^2-M^2\right)^2}+\frac{\left(m^2 M^2\right) \log \left(\frac{M^2}{m^2}\right)}{\left(m^2-M^2\right)^3}\right),\\
\Sigma'_{V+A}(m,M)=& -\frac{1}{8\pi^2}\left[\frac{1}{3} \log \left(\frac{\mu ^2}{m M}\right)+\frac{m^4-8 m^2 M^2+M^4}{9 \left(m^2-M^2\right)^2}\right.\notag\\
&\left.
+\frac{\left(m^2+M^2\right) \left(m^4-4 m^2 M^2+M^4\right) \log \left(\frac{M^2}{m^2}\right)}{6 \left(m^2-M^2\right)^3} \right].
\end{align}
We adopt dimensional regularization to regulate the UV-divergence, with $\mu$ the renormalization scale set at $\mu=M_Z$. But due to zeroth naturalness relation, the oblique parameters are free of divergence and therefore the choosing of scale $\mu$ is irrelevant. As a matter of fact, this feature provides a good way to check if the result is correct, which is important in particular for the models involving a couple of loop particles with mixing. 
 
For the above loop functions, there are two special cases that need to be handled with care during numerical processing. First is the degenerate case with $m=M$, then
\begin{align}
\Sigma_{V+A}&=  -\frac{1}{8\pi^2}m^2\textrm{ln}\frac{m^2}{\mu^2} , \quad 
\Sigma_{V-A}\text{=}  \frac{1}{8\pi^2}m^2\textrm{ln}\frac{m^2}{\mu^2},
\\
\Sigma'_{V+A}&=-\frac{1}{8\pi^2}\left(\frac{1}{3} \log \left(\frac{\mu ^2}{m^2}\right)-\frac{1}{6}\right), \quad 
\Sigma'_{V-A}\text{=}- \frac{1}{8\pi^2}\frac{1}{6}.
\end{align}
This case naturally occurs for the scalar functions of neutral vector bosons, $\Sigma_{ZZ}$ and $\Sigma_{3Y}$; it also arises for $\Sigma_{WW}$ when it receives contribution from the degenerate components of a $SU(2)_L$ multiplet. The second is the case with one massless particle, namely $M>m=0$, 
\begin{align}
\Sigma_{V+A}=&-\frac{1}{32\pi^2}M^2\left(2\textrm{ln}\frac{M^2}{\mu^2}-1\right),\quad 
\Sigma_{V-A}=0,\\
\Sigma'_{V+A}=&-\frac{1}{8\pi^2}\frac{1}{9} \left(1+3 \log\frac{\mu^2}{M^2}\right), \quad 
\Sigma'_{V-A}\text{=}0.
\end{align}
All of them are obtained by taking smooth limits of the general expressions~\footnote{The third case has both massless fermions, $M=m=0$, then $\Sigma_{V\pm A}$ vanish, but $\Sigma'_{V\pm A}$ are intermediate and require further treatment of the expression of parameters $S$, $T$ and $U$. But this case is not our concern.}.  


\subsection{Oblique parameters in the VLL-RHN  model}

In this subsection, we present the expressions of $S$, $T$ and $U$ specific to our simplified VLL-RHN model. We find a similar model setup in Ref.~\cite{Cynolter:2008ea}, which considers a doublet VLL mixing with a Dirac singlet fermion and presented the analytical results. Especially, Ref.~\cite{Cai:2016sjz}, which we only learned about in the final completion stage of this work, also studied doublet VLL mixing with Majorana fermions in the dark matter scenario. However, it is still of importance to make an independent calculation for cross check. But Ref.~\cite{Cai:2016sjz} adopt a different calculation procedure and make the direct comparison via expression difficult. In the numerical analysis, we will comment on this at the right place. 

With the loop functions $\Sigma_{V\pm A}$ and $\Sigma'_{V\pm A}$, and as well the current couplings, it is straightforward to obtain the total scalar functions like $\Sigma$ and $\Sigma'$ by summing over the loop diagrams. Then substitute them into Eq.~\eqref{STU parameterization}, we eventually get the oblique parameters
\begin{align}
S=&
-16\pi\left[\frac{1}{2}\tilde{\Sigma'}(m_L,m_L)-
4\sum_{a,b=1,}^{2,3}\right.\notag\\
&\left.\L\frac{|(g_L)_{ab}|^2+|(g_R)_{ab}|^2}{2}\Sigma_{V+A}'(M_a,M_b)+\text{Re}((g_L)_{ab}(g_R)_{ab}^*)\Sigma_{V-A}'(M_a,M_b)\R\right],
\end{align}
where the first and second term denote the contribution from the neutral currents of $e_4$ (with Dirac mass $m_L$) and neutral leptons (with Majorana masses $M_a$), respectively. In contrast, $T$ and $U$ receive contributions from charged currents,
\begin{align}
&T=\frac{4\pi}{M_Z^2s^2c^2}\left[\sum_{m=m_L}^{a=1,2,3}\L\frac{V_{1a}^2+V_{2a}^2}{{2}}\Sigma_{V+A}(m,M_a)+\text{Re}(V_{1a}V_{2a})\Sigma_{V-A}(m,M_a)\R 
-\frac{1}{2}\tilde{\Sigma}(m_L,m_L) \right.\notag\\&\left.
-4\sum_{a,b=1,}^{2,3}\L\frac{|(g_L)_{ab}|^2+|(g_R)_{ab}|^2}{2}\Sigma_{V+A}(M_a,M_b)+\text{Re}((g_L)_{ab}(g_R)_{ab}^*)\Sigma_{V-A}(M_a,M_b)\R\right],
\end{align}
and 
\begin{align}
&U= 16\pi\left[\sum_{m=m_L}^{a=1,2,3}\L\frac{V_{1a}^2+V_{2a}^2}{{2}}\Sigma'_{V+A}(m,M_a)+\text{Re}(V_{1a}V_{2a})\Sigma'_{V-A}(m,M_a)\R 
-\frac{1}{2}\tilde{\Sigma}(m_L,m_L) \right.\notag\\&\left.
-4\sum_{a,b=1,}^{2,3}\L\frac{|(g_L)_{ab}|^2+|(g_R)_{ab}|^2}{2}\Sigma_{V+A}(M_a,M_b)+\text{Re}((g_L)_{ab}(g_R)_{ab}^*)\Sigma_{V-A}(M_a,M_b)\R\right].
\end{align}
The above expressions can be easily generalized to the situation with more RHNs. 

We end up this section with a comment on the subtlety in calculating the contribution from a Majorana loop, and the above parameters are numerically finite only this subtlety is properly handled. For such a loop, in addition to the symmetry factor $\frac{1}{2}$, there is another $\frac{1}{2}$ from the Majorana nature in the couplings Eq.~\eqref{1/2factor}, which is cancelled by the factor 2 from the Feymann rule corresponding to the vertex Eq.~\eqref{NC}~\cite{Rosiek:1995kg}. From the example in Appendix.~\ref{cal:T}, one can track the difference between a Majorana and Dirac loop, which might provide a clue to distinguish the two scenarios. Overall, in previous studies, there has not been such detailed processing of the calculation of the oblique parameters in this model, which we believe is useful for researchers interested in this type of model.


\section{Constraints on the well-mixed VLL-RHN system}

In this section, we make the numerical study on the VLLs assisted with the seesaw (or, more generally,  the fermionic singlet-doublet model), taking advantage of the constraints from $S$ and $T$ before and after the CDF-II result. The key features of this model are captured by the VLL-RHN mass matrix. As a simplification, we will work in the minimal model with one family of RHN and VLL, which will contain three real parameters, $m_L$, $M_N$ and  $m_D$ if $m_D'$ is turned off. They are sufficient to investigate the general feature; the quantitative effect after including $m_D'$ in particular with a phase will be discussed separately.

\subsection{EWPO w/o CDF-II results}

To make precise predictions of EWPOs in new physics models, one follows the procedure of EW precision tests. In this procedure, the fine structure constant, Fermi constant and $Z$ boson pole mass are three most precisely measured EW quantities, and therefore are taken as fiducial quantities:  
\begin{equation}
\begin{split}
\hat \alpha(0)\approx 1/137.04,\quad \hat G_F\approx 1.164\times 10^{-5}\; {\rm GeV}^{-2}, \quad \hat M_Z \approx 91.1876\; {\rm GeV}.
\end{split}
\end{equation}
As a first step, three Lagrangian parameters $e, m_Z$ and $\sin^2\theta_w\equiv s^2$, which are used in calculating EWPOs both at tree and loop level, should be expressed in terms of the fiducial quantities (and the self-energy functions $\Sigma$). Then, the values of EWPOs, both ${\cal O}_{\rm SM}$ and the oblique parameters are functions of the fiducial quantities. For instance, the pole mass of $W$ boson takes the form of 
\begin{align}
M_W^2=
&{M}_Z^2{c}^2[1-\frac{c^2}{c^2-s^2}\frac{\delta\Pi_{ZZ}(M_Z^2)}{M_Z^2}+\frac{\delta\Pi_{WW}(M_W^2)}{M_W^2}+\frac{s^2}{c^2-s^2}(\delta\Pi'_{\gamma\gamma}(0)+\frac{\delta\Pi_{WW}(0)}{M_W^2}) ] \notag\\
=&{M}_W^2(\hat \alpha,\hat G_F,\hat M_Z)\left[1+\frac{\hat\alpha}{\hat c^2-\hat s^2}\L-\frac{1}{2}S+\hat c^2T+\frac{\hat c^2-\hat s^2}{4 \hat s^2}U\R\right]\label{W mass square shift=stu}.
\end{align}
with $\hat s^2 \approx 0.234$. The corresponding SM prediction ${M}_W^2(\hat \alpha,\hat G_F,\hat M_Z)\approx  80.3564\;\rm{GeV}. $

The $W$ boson mass is an attractive EWPO to probe new physics, since it is not sensitive to the strong interaction and has amazingly small uncertainty in the SM theoretical prediction. This value is obtained from indirect determination by global fit with uncertainty $\sim 0.01\%$ (corresponding to $\delta M_W \lesssim 10$ MeV); see the improved predictions with respect to different colliders in the last column  of  \tabref{tab:my_label}. Therefore, as long as the accuracy of directly measuring $W$ boson mass can reach this level, we can look for possible new physical hints from this EWPO~\footnote{In contrast, the indirect determination of Higgs and top masses suffer a relatively large theoretical uncertainty, which hinders them from becoming EWPOs sensitive to new physics.}, and the third column of \tabref{tab:my_label} shows the current situation. 

We are already probing the new physics domain. Of great interest is the CDF Collaboration at Fermilab, who published the most precise measurement of  $M_W$, analyzing the full dataset of the Tevatron collider. They reported a value of 80434 MeV and an uncertainty of 9 MeV, which differs significantly from the SM prediction, but also differs significantly from the other experimental results. In a new preliminary result released by the ATLAS Collaboration, with an improved re-analysis of its initial $M_W$ measurement, they found $M_W$ to be $80360\;\text{MeV}$, with an uncertainty of just 16 MeV, still in agreement with the SM. Our strategy in the face of such chaotic measurement results is to prepare with both hands, using global fit with/without CDF-II data. This was already carried out in Ref.~\cite{Lu:2022bgw}, and we quote their results for our analysis. In our model, both $S$ and $U$ will be found to be small, and we only need to consider the constraint from $T$, at 2$\sigma$ confidence level, which gives
\begin{align}
  {\rm PDG-2021}: -0.010\,819\leq T\leq 0.116\,374,\quad  {\rm CDF-II}: 0.122\,222\leq T\leq 0.192\,398.
\end{align}
In any case, the current sensitivity to $T$ is $\sim {\cal O}(0.1)$. For the latter, the $W$-boson mass anomaly solely determines $T\simeq 0.1$. 
\begin{table}[htbp]
    \centering
    \begin{tabular}{|c|c|c||c|}
         \hline
         Colliders&experiments&results&SM Prediction\\
         \hline
         \multirow{2}{*}{LEP}&LEP&$80440 \pm 43(stat.)\;\text{MeV}$~\cite{ALEPH:2006cdc}&$80373\pm 23\;\text{MeV}$~\cite{ALEPH:2005ab}\\
         ~ & LEP combination~\cite{ALEPH:2013dgf} &$80376\pm33\;\text{MeV}$ & $80385\pm 15\;\text{MeV}$\\
         \hline
         \multirow{2}{*}{Tevatron}& D0 (Run 2)~\cite{D0:2009yxq,D0:2012kms} & $80375\pm23\;\text{MeV}$ & $80399\pm23\;\text{MeV}$~\cite{ALEPH:2010aa} \\
         & CDF (Run 2)~\cite{Hays:2022qlw} & $80433.5\pm9.4\;\text{MeV}$ & $80357\pm6\;\text{MeV}$\\
         \hline
         \multirow{4}{*}{LHC}& LHCb 2022~\cite{LHCb:2021bjt} & $80354\pm23(stat.)\;\text{MeV}$ &$80379\pm12\;\text{MeV}$~\cite{ParticleDataGroup:2020ssz} \\
         ~& ATLAS 2017~\cite{ATLAS:2017rzl} & $80370 \pm 19\;\text{MeV}$ &$80385 \pm 15\;\text{MeV}$~\cite{ParticleDataGroup:2014cgo} \\
         ~& ATLAS 2023~\cite{ATLAS:2023fsi} & $80360 \pm 16\;\text{MeV}$ &$80377 \pm 12\;\text{MeV}$~\cite{ParticleDataGroup:2022pth} \\
         ~& ATLAS 2024~\cite{ATLAS:2024erm} & $80366.5 \pm 15.9\;\text{MeV}$ & $80355 \pm 6\;\text{MeV}$~\cite{deBlas:2021wap}\\
         \hline
    \end{tabular}
    \caption{Overview of $W$ boson mass.}
    \label{tab:my_label}
\end{table}


\subsection{The minimal VLL-RHN system with \texorpdfstring{$m_D'\to 0$}{}}

This minimal case with one RHN and moreover $m_D'\to 0$ corresponds to the gauged $(B-L)_{ij}$ model with one decoupled RHN. We will see that, the VLL can only leave a significant imprint in the $T$ parameter, in the well-mixed region of VLL-RHN that leads to a large custodial symmetry breaking. We have a vanishingly small $U$ as usual, and $S$ is also suppressed. Actually, although not related to custodial symmetry breaking, like $T$, $S$ also vanishes in the limit of vanishing doublet-singlet mixing (similar observation is made in other models~\cite{Cao:2022mif}), since it recovers $SU(2)_L$ . Hence, the mass mixing parameter $m_D$ or $\lambda_n$ is a key to enhance the oblique parameters.

\begin{figure}[H]
    \centering
    {
    \includegraphics[width=0.3\textwidth]{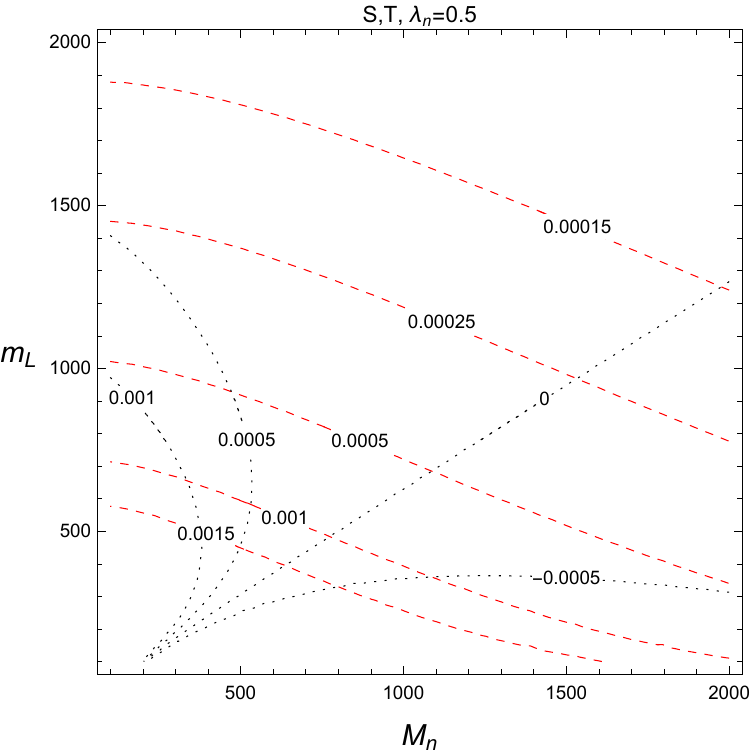}
    }
    {
    \includegraphics[width=0.3\textwidth]{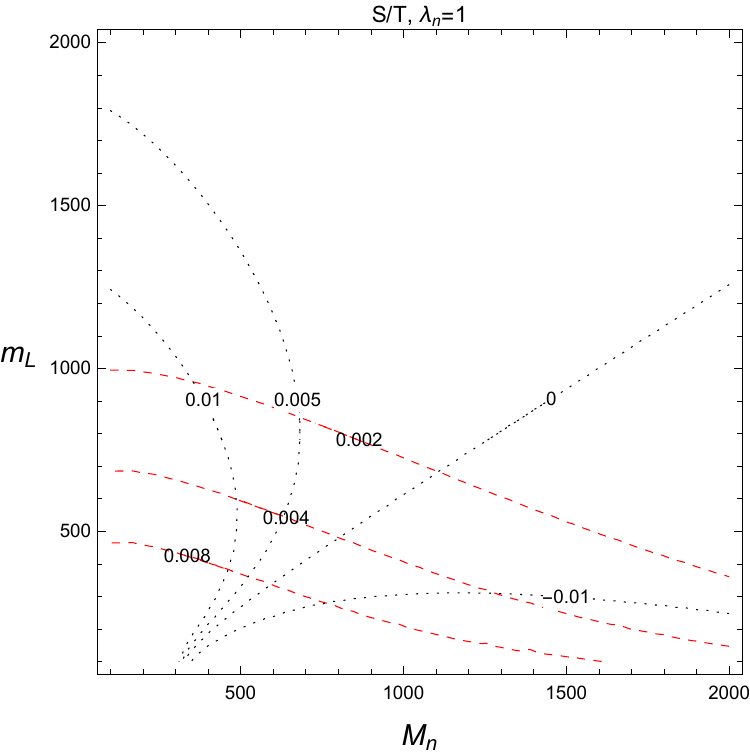}
    }
    {
    \includegraphics[width=0.3\textwidth]{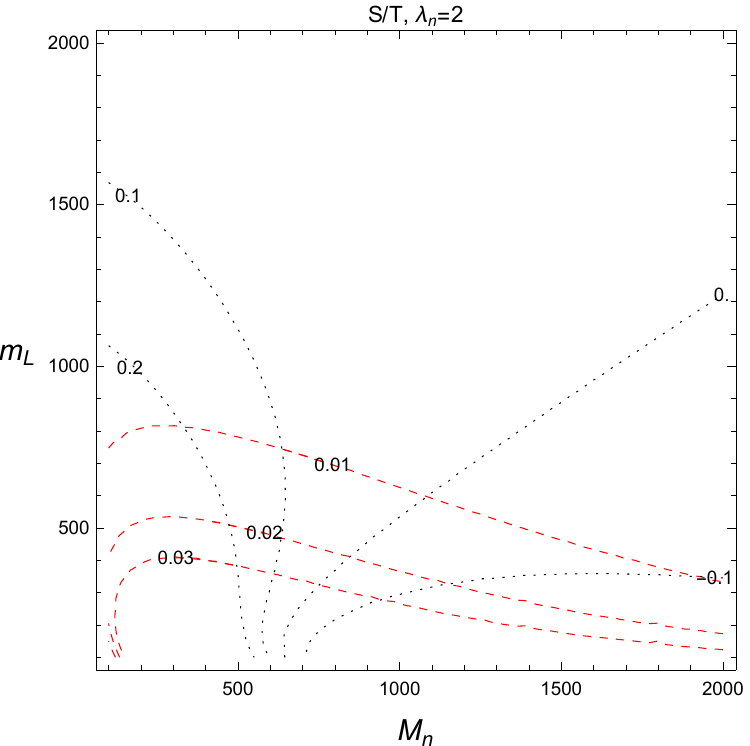}
    }
    \caption{The distributions of parameters $S$ (red-dashed) and $T$ (black-dotted) in the $M_N - m_L$ plane for three typical points - $\lambda_n = 0.5 ~(\rm left) ,\, 1 ~ (\rm middle) ,\, 2~ (\rm right)$.}
    \label{fig.yukawaST}
\end{figure}

To explore the overall features of the three dimensional parameter space, we fix $M_N$ at three typical scales, a sub weak scale 10 GeV, weak scale 100 GeV and the TeV scale, and then plot the oblique parameters on the $\lambda_n-m_L$ plane. The VLL mass lies in the region $100{~\rm GeV}<m_L<2000{~\rm GeV}$, where the lower bound is due to constraint on the charged heavy lepton mass, from the robust LEP-II bound~\cite{L3:2001xsz}, and the precise lower bound does not matter much in our discussion. Moreover, we limit $0<\lambda_n<3$, with the upper bound imposed simply by hand to avoid hitting the Landau pole at a fairly low scale. The region with $-3<\lambda_n<0$ is symmetric with the previous one thus not considered. We present some plots and make some observations: 
\begin{enumerate}
    \item The $S$ parameter is shown in the top panels of Fig.~\ref{fig.demoorderS/T}, and we see that its value always keeps very small,  $\lesssim {\cal O}(0.01)$. Such small values lie far below the sensitivity of the current experiments, $\gtrsim 0.1$. The value of $U$ parameter is even much smaller and not displayed. But the $T$ parameter can be sizable, except for a quite heavy RHN of several TeV; see the bottom panels of Fig.~\ref{fig.demoorderS/T}. 
  \begin{figure}[H]
    \centering
    {
    \includegraphics[width=0.3\textwidth]{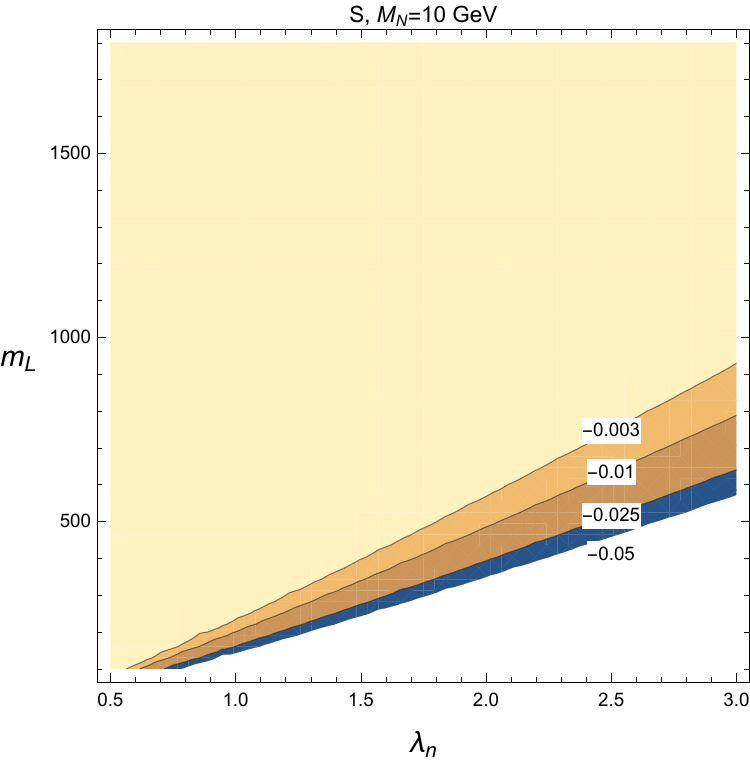}
    }
    {
    \includegraphics[width=0.3\textwidth]{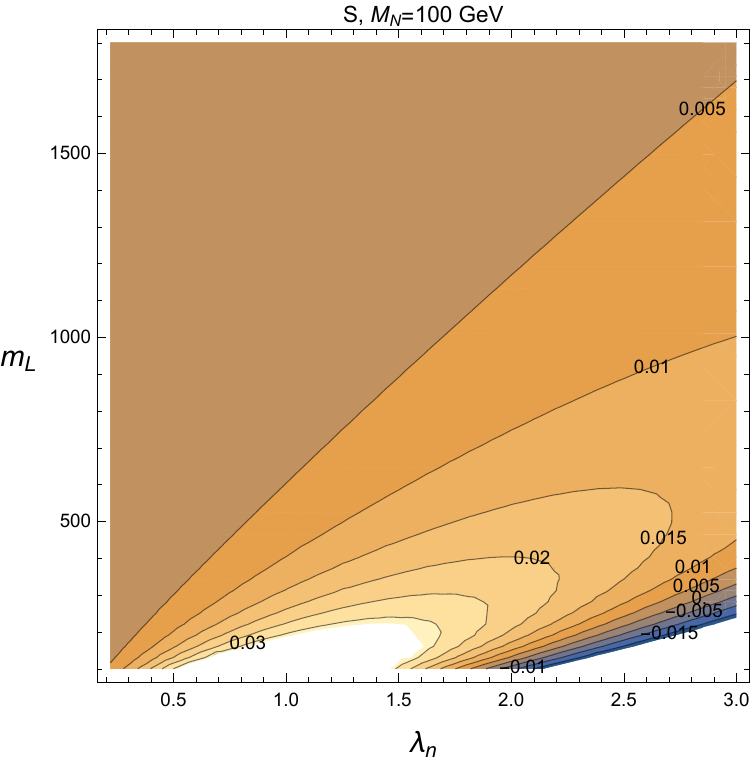}
    }
    {
    \includegraphics[width=0.3\textwidth]{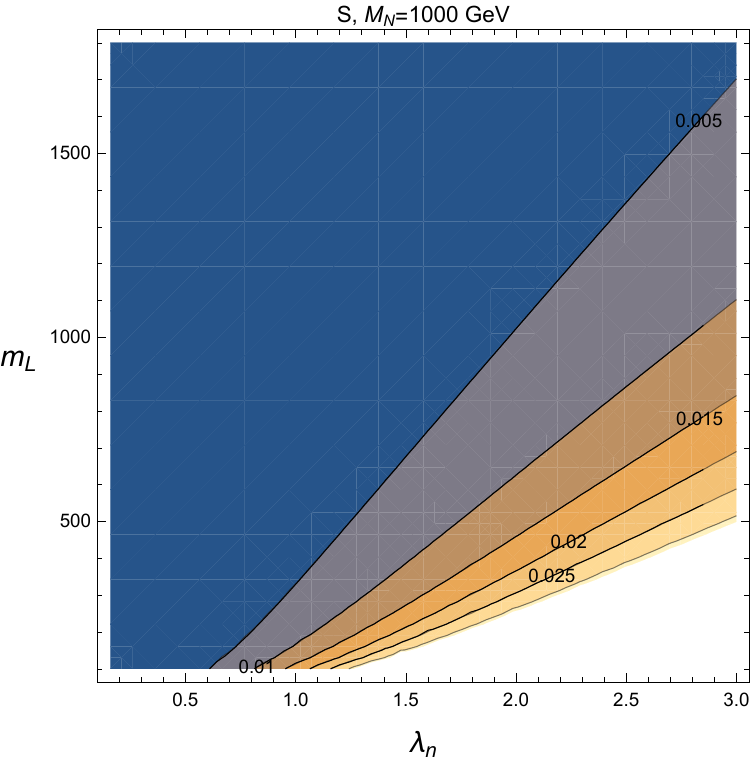}
    }
    \\
    {
    \includegraphics[width=0.3\textwidth]{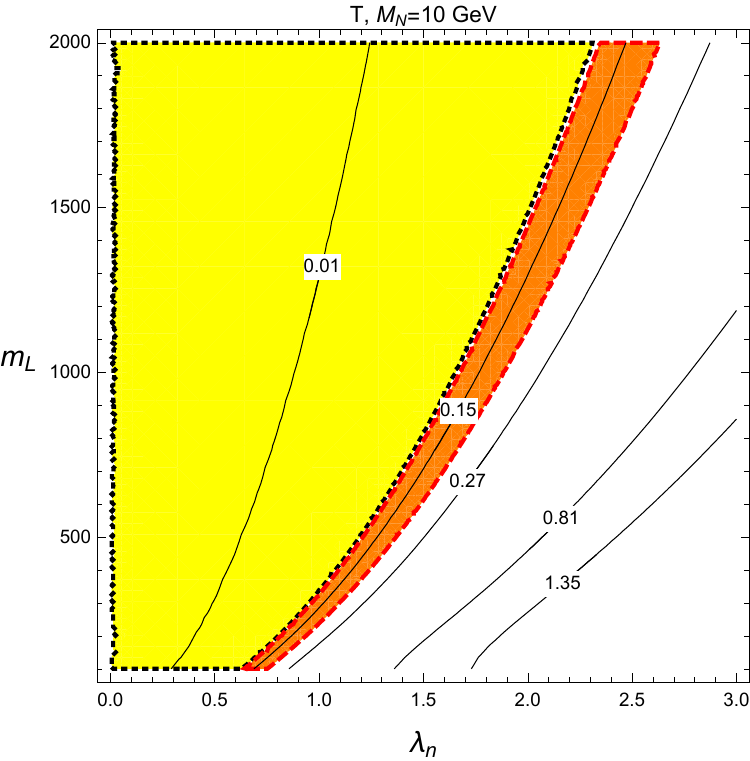}
    }
    {
    \includegraphics[width=0.3\textwidth]{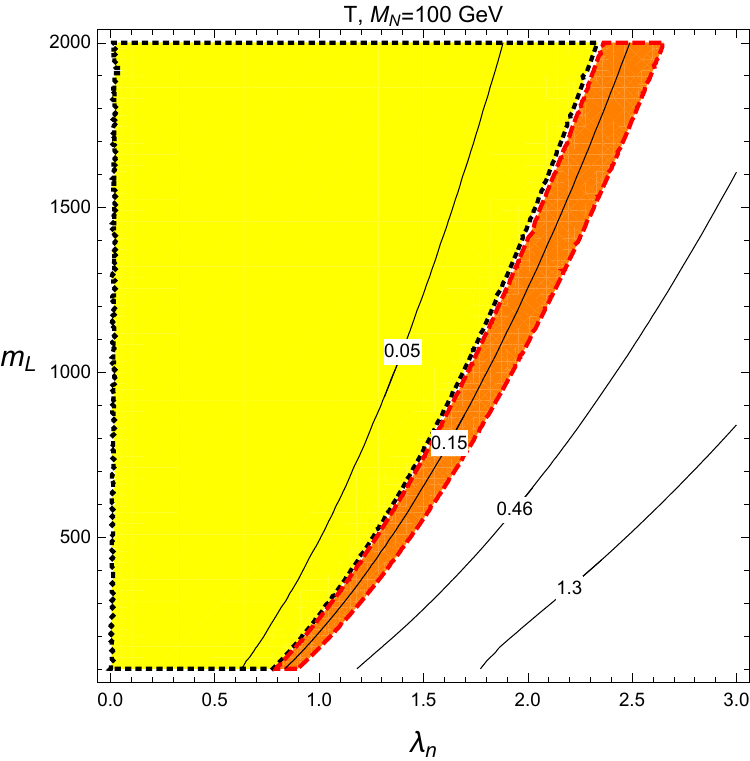}
    }
    {
    \includegraphics[width=0.3\textwidth]{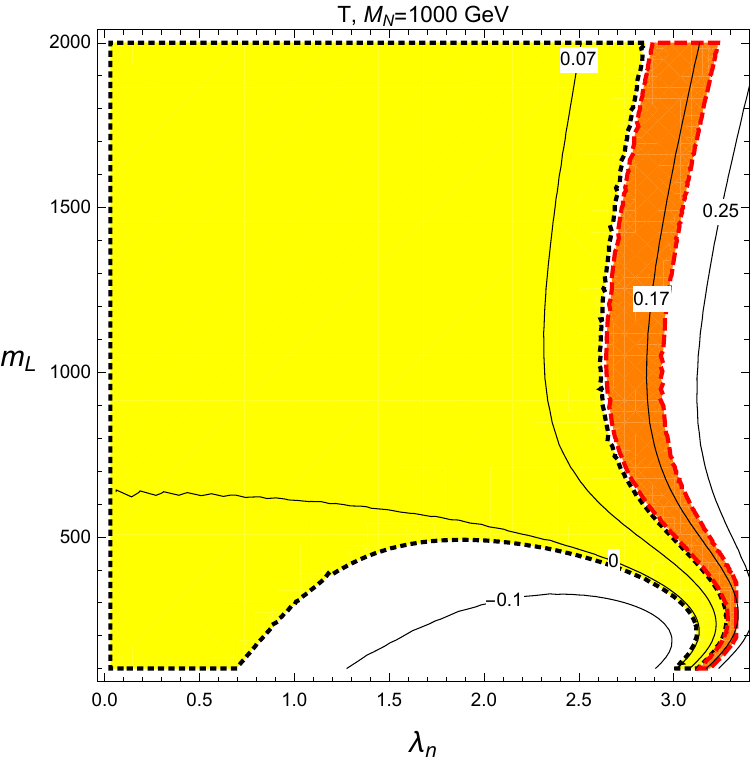}
    }
    \caption{The analyse of the order of magnitude of parameters between parameters $S$ (top) and $T$ (bottom) are showed, setting three typical reference points at $M_N=10^{3}\;\text{GeV}$, $M_N=10^{2}\;\text{GeV}$ and $M_N=10\;\text{GeV}$. In the bottom plots, we also show the allowed-region of $T$ by PDG-2021 (yellow shadowed) and CDF-II (orange shadowed) respectively.}
    \label{fig.demoorderS/T}
\end{figure}
    \item   It is well expected that for a fairly light RHN with $M_N\ll m_L$, the model reduces to a two parameters case (it explains why we do not consider the even lighter $M_N$ case), and this is in accordance with the strong similarity between plots for the $M_N=100$ GeV and $M_N=10$ GeV cases, in particular in the heavy $m_L$ region.
    \item Increasing the RHN mass causes the oblique parameters quickly go beyond the current sensitivity, because the VLL approximately decouples with RHN thus the restoration of custodial symmetry, except that the splitting is compensated by a very large $\lambda_n$. The decoupling behavior of new particles is reflected in the fact that, for a given $\lambda_n$, both $S$ and $T$ monotonically decrease with the increasing $m_L$, except for certain subtle regions where the strong mixing effect may break the simple decoupling behavior.  
   
    \item For $T$, across certain line, sign flipping may occur. It is found that in our choice of $M_N$, we meet this flipping  only in the case with $M_N=1$ TeV. This is related to the presence of Majorana particles, which, contrary to the Dirac fermions, have  non-diagonal couplings with $Z$ and thus contribute negatively to its self energy. 
    
Actually, a negative $T$ is not rare in the parameter space. To show this, let us instead display the plots of $S$ and $T$ in the $M_N-m_L$ plane with fixed $\lambda_n$, see  \figref{fig.yukawaST}. They clearly show that there is a (almost) straight line along which one meets the accidental cancellation leading to $T=0$, below which $T<0$. A negative $T$ is not the focus of this work, and we refer to Ref.~\cite{Ma:1992uc} for a deep understanding the origin of such $T$. 
\end{enumerate}

It is of interest to use this to constrain on the mixed doublet-singlet system. The corresponding regions are shown in the bottom panels of Fig.~\ref{fig.demoorderS/T}, shadowed with yellow and orange colors, respectively; they do not have overlap. In the following we will take the CDF-II region (orange) as a smoking gun for the coming hint of the VLL-RHN system in the oblique parameter. For the relatively light $m_N$, the CDF-II  region can be accommodated with a moderately large $\lambda_n\sim 1$ even for a TeV scale $m_L$. But for $m_N=1$ TeV, a fairly large $\lambda_n\approx 3$ is needed. In other words, with current sensitivity, our proposal is hopeful only for the sub-TeV scale RHN.

\subsection{Phase effect after including  \texorpdfstring{$m_D'$}{}}

\begin{figure}
    \centering
    \subfloat[]
    {
    \includegraphics[width=0.45\textwidth]{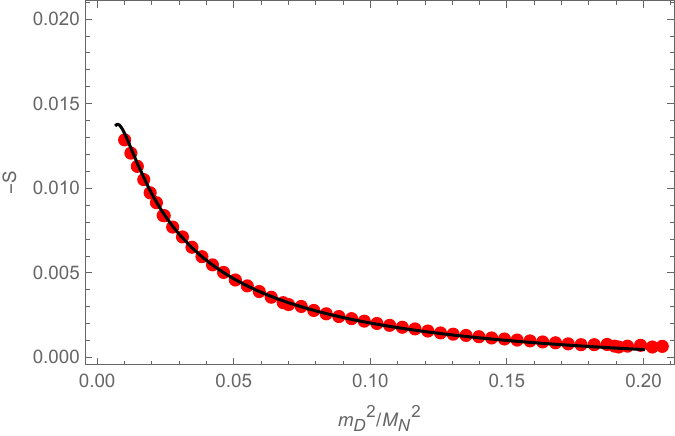}
    }
    \subfloat[]
    {
    \includegraphics[width=0.45\textwidth]{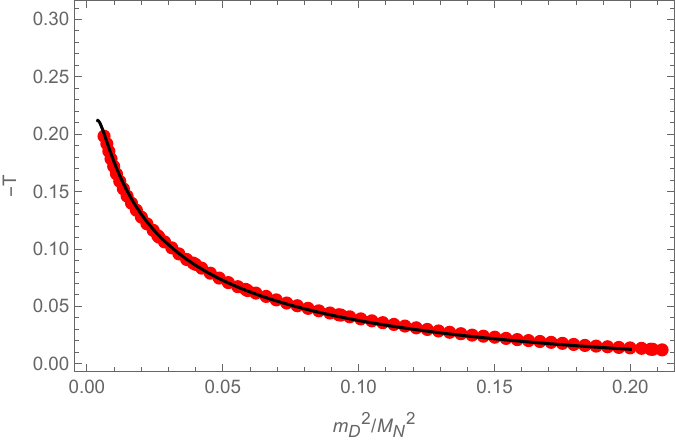}
    }
    \caption{Comparison of the $S$ and $T$ parameters between our calculation (black-solid lines) and the results (red-dotted lines) quoted from Fig.~1 and Fig.~2 of ~Ref.~\cite{Ma:1992uc}, taking the same inputs.}
    \label{fig.Test1}
\end{figure}

We now go beyond the minimal case, turning on both $m_D$ and $m_D'$. This case coincidences with the one studied in Ref.~\cite{Ma:1992uc}, so it is a good place to compare our result with theirs and we have confirmed complete consistency taking the same parameters, shown in~\figref{fig.Test1}~\footnote{
Moreover, we confirmed that if $m_D=\pm m_D'$, $T$ and $U$ vanish as observed in ~\cite{Cai:2016sjz}. But we are not sure if $\left|m_D\right| = \left|m_D'\right| \not = 0$ is related to the usual custodial symmetry.}. 

It is of interest to investigate the possible phase effect which may arise in this case~\footnote{It also appears in cases with more RHNs.} but not discussed before. Now, the mass matrix contains four complex elements $M_N$, $m_L$, $m_D'$ and $m_D$, eight real degrees of freedom. But their phases can be absorbed via re-definition of fields, leaving only one physical phase $\theta$. We explicitly show this manipulation via the phase rotation $P={\rm Diag}(e^{-i(\beta-\theta_1+\alpha/2)},e^{-i(\theta_1-\alpha/2)},e^{-i\alpha/2})$:
\begin{align}
M_{\psi}=&
\begin{pmatrix}
    0 & m_L e^{i \beta} & m_D e^{i \theta_2} \\[0.5ex]
    m_L e^{i \beta} & 0 & e^{i \theta_1}  m_D'  \\[0.5ex]
    m_D e^{i \theta_2} & e^{i \theta_1} m_D' & M_N e^{i \alpha}
\end{pmatrix} \to 
P^T M_{\psi} P =
\begin{pmatrix}
     0 & m_L & m_D e^{i \theta} \\[0.5ex]
     m_L & 0 & m_D'  \\[0.5ex]
    m_D e^{i \theta} & m_D' & M_N
\end{pmatrix}.
\end{align}
where $\theta = \theta_1 + \theta_2 - \alpha - \beta$. Thus we only consider the phase of $m_D$ here. This phase may give rise to a significant effect in some parameter space, as discussed below. 

\begin{figure}[H]
    \centering
    {
    \includegraphics[width=0.45\textwidth]{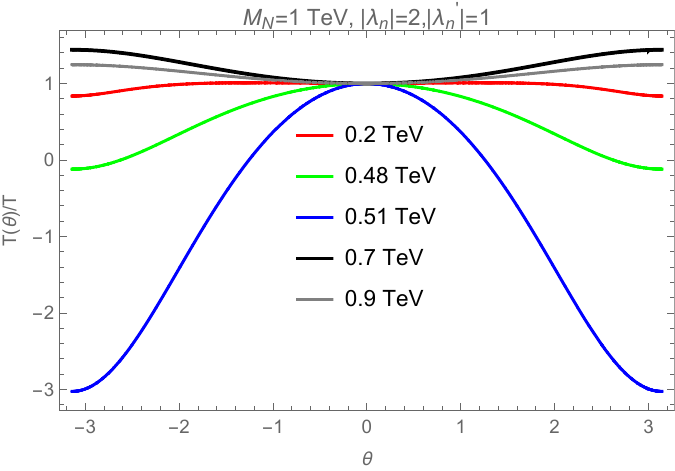}
    }
    {
    \includegraphics[width=0.47\textwidth]{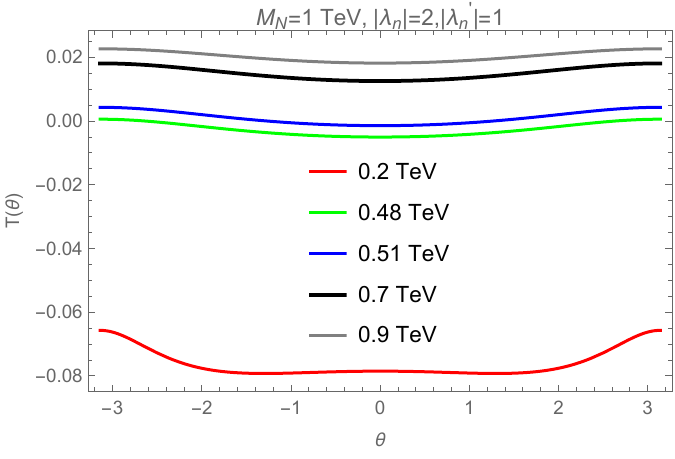}
    }
    \caption{The figures show the effect of phase with $|\lambda_n| = 2$, $|\lambda'_n| = 1$, $M_N =1 \; {\rm TeV}$, taking $m_L = 0.2, \, 0.48, \, 0.51, \, 0.7, \, 0.9 \; {\rm TeV}$ labeled as the red, green, blue, black, gray lines respectively. $T(\theta)/T(0)$ for the left panel and $T(\theta)$ for the right panel.}
    \label{fig.Tphase3}
\end{figure}

Following Eq.~\eqref{CC}, it is clear that the effect of the possible phase exhibit itself via the chiral mixing terms in the charged current, such as the $\bar{N}_{a}\gamma^{\mu}e_4$ contribution 
\begin{equation}\label{CC:a4}
    \frac{g}{\sqrt{2}} \left(V_{2 a}^* P_L + V_{1 a} P_R\right) \to \frac{g^2}{2} {\rm Re}
    \left(V_{2 a} V_{1 a}\right) \, , \quad a=1,2,3 \, .
\end{equation}
We find that in the region which has a large mixing between VLL and RHN, the $\theta$ phase effect can play a non negligible role. To show this, in \figref{fig.Tphase3} we plot $T(\theta)/T(0)$ and $T(\theta)$ for the example with $M_N=1.0$ TeV, $|\lambda_n|=2.0$ and  $|\lambda_n'|=1.0$; we take several values of $m_L$, and find that for the sake of a large phase effect, $m_L$ should not exceed $M_N$ when yukawa couplings are about order one. Varying the phase $\theta$ may cause a considerably increased or decreased $T(\theta)/T(\theta=0)$; in particular, a vanishing $T(\theta)$ may be realized, which is due to a subtle and complicated cancellation by tuning the parameters. For this example, it happens around $m_L=0.5$ TeV and $m_D\simeq m_L$, $m_D' \simeq m_L$ where Eq.\eqref{CC:a4} are about order one. However, such region is related to the vanishing of $T$, and therefore if it is of practical interest is questionable.

\section{Conclusion and discussion}

Hopefully, the ${\rm SM}_{\nu,I}$, the type-I seesaw mechanism extension to SM, is going to be the next SM. Nevertheless, testing the prediction of this model is hampered by the highly decoupling between the Majorana RHNs and SM. In this work, we consider a simple scenario where VLLs, which are introduced in different contexts, enter the ${\rm SM}_{\nu,I}$ via the RHN portal. In the well-mixed VLL-RHN region, there are many interesting features which help to discover such a system.

As a preliminary study, in this work we analytically calculate the Peskin-Takeuchi parameters. The calculation is routine, but the Majorana loop should be treated carefully in a system with complicated mixings, otherwise it is impossible to obtain a correct result. We checked our expressions by examining several points, such as  free of divergence, vanishing numerically in special limits which are related to symmetry recovery. The VLL-RHN system only gives a sizable $T$ parameter, and the current sensitivity is  $T\lesssim{\cal O}(0.1)$, which probes RHNs and VLLs below the TeV scale, with a properly large mixing. In particular, the $W$ boson mass reported by the CDF-II Collaboration can be readily accommodated, but we need more time to see this if this anomaly survives. 

After this work, we will continuously to study the other aspects of the well-mixed VLL-RHN system. The region related to the CDF-II  result is of special interest, since it points to new electroweak fermions not far from the TeV scale, which are within the LHC reach.




\newpage

\section{Appendix}

\subsection{Scalar functions in different basis}\label{ps.various expressions}

For the electroweak part of SM, the gauge interactions before electroweak spontaneously breaking take the forms of 
\begin{equation}
\begin{split}
\mathcal{L}_{EW}=& g(W_1^\mu J^1_\mu+W_2^\mu J^2_\mu)+gW_\mu^{3}J_\mu^{3}+g'B_\mu J_\mu^{Y},
\end{split}
\end{equation}
with the currents defined as 
\begin{equation}\label{current1}
    \begin{split}
J^{1,2,3}_{\mu} &= \sum_i \bar{\psi}_{Li}T^{1,2,3}\gamma_{\mu}\psi_{Li},\quad 
J^{Y}_{\mu} = \sum_i (\bar{\psi}_{Li}\frac{Y}{2}\gamma_{\mu}\psi_{Li}+\bar{\psi}_{Ri}\frac{Y}{2}\gamma_{\mu}\psi_{Ri}).
    \end{split}
\end{equation}
where $T^{1,2,3}$ are three generators of $SU(2)_{L}$, $Y$ is the hypercharge of $U(1)_{Y}$. Contributions to the currents from new particles can be easily incorporated. In the basis after electroweak spontaneously breaking, it take the forms of 
\begin{equation}
\begin{split}
\mathcal{L}_{EW}
=&\frac{g}{\sqrt{2}}(W^{+\mu} J^+_\mu+W^{-\mu} J^-_\mu)+gZ^\mu J^Z_\mu+eA^\mu J^{A}_\mu.
\end{split}
\end{equation}
The notations $J_\mu^{A}=J_\mu^{EM}=e\,J_\mu^{Q}$ also appear in some literature. The above currents are the linear combinations of those currents in Eq.~(\ref{current1})
\begin{equation}\label{relation:current}
    \begin{split}
J^{\pm}_{\mu} &= J^{1}_{\mu} \pm i J^{2}_{\mu} = 
 \sum_i \bar{\ell}_{Li}T^{\pm}\gamma_{\mu}\ell_{Li},\\
 J^{Z}_{\mu} &=  \frac{1}{\textrm{c}_{\textrm{W}}}(J^{3}_{\mu} - \textrm{s}_{\textrm{W}}^2J^{EM}_{\mu}) = \sum_i \frac{1}{\textrm{c}_{\textrm{W}}}\bar{\psi}_{i}(T^{3}P_L-\textrm{s}_{\textrm{W}}^2Q)\gamma_{\mu}\psi_{i},\\
J^{EM}_{\mu} &= \sum_i Q_{i}(\bar{\psi}_{Li}\gamma_{\mu}\psi_{Li}+\bar{\psi}_{Ri}\gamma_{\mu}\psi_{Ri}),
    \end{split}
\end{equation}
with the charge $Q=T^3+\frac{Y}{2}$.  As authors' preferences in the original reference~\cite{Peskin:1991sw,Peskin:1990zt}, $J^Z_\mu$ is expressed in terms of $J_\mu^3$ and $J_\mu^{EM}$. This will lead to the appearance of scalar functions in the basis before electroweak symmetry breaking, such as $\Sigma_{QQ,3Q}$ in  Eq.~(\ref{relation1}).


The vacuum polarization amplitude between two vector boson $V$ and $V'$, $\Sigma_{VV'}$ are written in the momentum space as  
\begin{equation}
    \begin{split}\label{defination of vacuum polarization amplitude}
&\Sigma^{\mu\nu}(q^2)=\int\mathrm{d}^{4}x\;\exp(-iqx)\langle J^{\mu}(x)J^{\nu}(0)\rangle,
    \end{split}
\end{equation}
which can be decomposed into the transverse and longitudinal parts 
\begin{equation}
i\Sigma_{VV'}^{\mu\nu}(q^2)=P^{\mu\nu}i\Sigma_{VV'}^{T}(q^2)+L^{\mu\nu}i\Sigma_{VV'}^{L}(q^2),
\end{equation}
with the transverse and longitude projection operators defined as
\begin{equation}
    \begin{split}
P^{\mu\nu}=g_{\mu\nu}-\frac{q_\mu q_\nu}{q^2},\quad L^{\mu\nu}=\frac{q_\mu q_\nu}{q^2} ,\quad P^{\mu\nu}+L^{\mu\nu}=g_{\mu\nu},\\
P_{\mu\nu}P^{\mu\nu}=3,\quad L_{\mu\nu}P^{\mu\nu}=0,\quad P_{\mu\nu}L^{\mu\nu}=0,\quad L_{\mu\nu}L^{\mu\nu}=1.
    \end{split}
\end{equation}
The scalar function for the transverse part is obtained $\Sigma_{VV'}^{T}(q^2)=\frac{1}{3}P_{\mu\nu}\Sigma_{VV'}^{\mu\nu}(q^2)$. It corresponds to $-i$ times the loop diagram by Peskin's convention. 
One can use the above notation to rewrite the expression of the scalar function for $W$, $Z$, and $\gamma$ as 
\begin{equation}
    \begin{split}\label{relation1}
&\Sigma_{\gamma\gamma}=e^2\Sigma_{QQ},\\
&\Sigma_{ZZ}=\frac{g^2}{c_W^2}\Sigma_{33},\\
&\Sigma_{ZA}=\frac{g}{c_W}e(\Sigma_{3Q}-s^2\Sigma_{QQ}),\\
&\Sigma_{WW}= g^2\Sigma_{11}+g^2\Sigma_{22}= 2g^2\Sigma_{11}=2g^2\Sigma_{22}.
    \end{split}
\end{equation}
It's self-evident that these different expressions of oblique parameters are equivalent.

If the energy scale of the new physics is heavy enough, then the derivative in the oblique parameters can be written as
\begin{equation}
    \begin{split}
        &\Sigma_{33}'(0)=\frac{\Sigma_{33}(M_Z^2)-\Sigma_{33}(0)}{M_Z^2},\quad
        \Sigma_{33}'(0)=\frac{\Sigma_{33}(M_W^2)-\Sigma_{33}(0)}{c^2M_Z^2},\\
        &\Sigma_{3Q}'(0)=\frac{\Sigma_{3Q}(M_Z^2)-\Sigma_{33}(0)}{M_Z^2},\quad
        \Sigma_{3Y}'(0)=\frac{\Sigma_{3Y}(M_Z^2)-\Sigma_{3Y}(0)}{M_Z^2}.
    \end{split}
\end{equation}
Here we ignored $\Sigma_{QQ}$ since it is trivial, and in some special-chosen models, $\Sigma'_{VV'}$ and $\Sigma''_{VV'}$ would be needed too.

\subsection{FCNCs of Majorana fields}\label{Majorana:interaction}

In this appendix, we give the details of how to rewrite the neutral current couplings involving four-component Majorana fields with flavor changing (i.e., FCNC) in the form of Eq.~(\ref{NC}). First, note that the Majorana bilinear terms satisfy the charge conjugate relations $\overline{\chi^C}\gamma^\mu\eta^C=-\bar{\eta}\gamma^\mu\chi$ or $\overline{\xi^C}\gamma^\mu P_L\zeta^C=-\bar{\zeta}\gamma^\mu P_R\xi$. So, for the original FCNC couplings in Eq.~(\ref{NC:1}) we have 
\begin{align}
    \mathcal{L}\supset&\frac{g}{{2}c}Z_{\mu}(\bar{N}_a\gamma^{\mu}V_{1a}V_{1b}^*N_b +\overline{N^C}_a\gamma^{\mu}V_{2a}^*V_{2b}N^C_b)\notag\\
    =&
    -\frac{g}{{2}c}Z_{\mu}(\bar{N}_a\gamma^{\mu}V_{2a}V_{2b}^*N_b +\overline{N^C}_a\gamma^{\mu}V_{1a}^*V_{1b}N^C_b)\notag\\
    =&
    \frac{g}{4c}Z_{\mu}[\bar{N}_a\gamma^{\mu}( V_{1a}V_{1b}^*-V_{2a}V_{2b}^* )N_b +\overline{N^C}_a\gamma^{\mu}( V_{2a}^*V_{2b}-V_{1a}^*V_{1b} )N^C_b].
\end{align} 
Using $P_L+P_R=1$, the above equation can be rewritten as
\begin{align}  
    &\frac{g}{4c}Z_{\mu}[\bar{N}_a\gamma^{\mu}V_{1a}V_{1b}^*N_b-\bar{N}_a\gamma^{\mu}V_{2a}V_{2b}^*N_b - \overline{N^C}_a\gamma^{\mu}V_{1a}^*V_{1b}N^C_b+\overline{N^C}_a\gamma^{\mu}V_{2a}^*V_{2b}N^C_b]\notag\\
    &=
    \frac{g}{4c}Z_{\mu}\bar{N}_a\gamma^{\mu}P_L(V_{1a}V_{1b}^*-V_{2a}V_{2b}^*)N_b -\frac{g}{4c}Z_{\mu}\overline{N}_a\gamma^{\mu}P_R(V_{1a}^*V_{1b}-V_{2a}^*V_{2b})N_b.
\end{align}
The coupling of left-handed and right-handed are interrelated through 
$g^Z_L=(g^Z_R)^*$, and this interrelation makes one additional factor $\frac{1}{2}$ appear.

 

We would like to comment that, FCNC is very small within the $\rm SM_{\nu,I}$, but it can be enhanced after introducing VLLs like in our setup. Then, the non-diagonal coupling between $N_1$ and $N_2$  source the negative contribution to parameter $T$. However, this negative sign is not tied to the Majorana property. For instance, if the mixing is between two charged components in the doublets with different hypercharge $T^3$, one can also obtain a negative $T$.


\subsection{A demo calculation}\label{cal:T}

In this appendix, we elucidate the complete process of calculating oblique parameters through a demo of parameter $T$. 

The $W^3$ contribution to parameter $T$ has two types of Feynman diagrams: the neutral fermion loop $\Sigma(M_a, M_b)$ and charged fermion loop $\Sigma(m_L, m_L)$. However, the neutral fermion is Majorana one here, and the Feynman rules differ from the Dirac fermion.
\begin{figure}[H]  
\centering    
\begin{minipage}{0.4\linewidth}
    \begin{tikzpicture}
    \begin{feynhand}
    \vertex (a) at (-5,0) ;
    \vertex (b) at (-3,0) ;
    \vertex (d) at (-1,0) ;
    \vertex (c) at (-3,2) ;
    \propag  [boson] (c) to (b);
    \propag  [fermion] (a) to (b);
    \propag  [fermion] (b) to (d);
    \node at (-5.5,0) {$N_a$};
    \node at (-0.5,0) {$N_b$};
    \node at (-3,2.5) {$W^3$};
    \node at (-3,-0.5) {$2g(g_L)_{ab}P_L+2g(g_R)_{ab}P_R$};
\end{feynhand}
    \end{tikzpicture}
\end{minipage}
\begin{minipage}{0.4\linewidth}
    \begin{tikzpicture}
    \begin{feynhand}
    \vertex (a) at (-5,0) ;
    \vertex (b) at (-3,0) ;
    \vertex (d) at (-1,0) ;
    \vertex (c) at (-3,2) ;
    \propag  [boson] (c) to (b);
    \propag  [fermion] (a) to (b);
    \propag  [fermion] (b) to (d);
    \node at (-5.5,0) {$e_4$};
    \node at (-0.5,0) {$e_4$};
    \node at (-3,2.5) {$W^3$};
    \node at (-3,-0.5) {g/2};
\end{feynhand}
    \end{tikzpicture}
\end{minipage}
    \caption{The corresponding Feynman rules of $W^3$ vertex: The Majorana-induced $\frac{1}{2}$ factor (the left) is compensated by two possible contractions compared with the right one. However one more $\frac{1}{2}$ appear subsequently as a symmetry factor at the expense of Majorana.}
    \label{twotypevertex}
\end{figure}

The $W^3$ gauge interaction contributes to parameter $T$ through $\Sigma_{33}$ which can be computed by Eq.~\eqref{new basis}, Eq.~\eqref{STU parameterization}, Eq.~\eqref{sigmamassandcoupling} and Eq.~\eqref{vector-typesigma}. The corresponding the mass variables are $M_a, M_b=m_L$, and the  coupling variables are $a=b=\frac{1}{2}$ \footnote{Follow the convention, $g$ and $g'$ don't appear in Eq.~\eqref{STU parameterization}. For Majorana loop,The vertex double its coupling-$(g_{L/R})_{ab}\to 2(g_{L/R})_{ab}$ and loop symmetry adds an additional $\frac{1}{2}$ factor. Here the index label $a$ or $b$ are $1,2,3$.}. Thus loop contribution from $e_4$ gives
\begin{align}
&\left[\left(\frac{1}{2}\right)^2+\left(\frac{1}{2}\right)^2\right]\Sigma_{V+A}(m_L,m_L)+2\cdot\text{Re}\left(\frac{1}{2}\cdot\frac{1}{2}\right)\Sigma_{V-A}(m_L,m_L) \nonumber\\
=&\frac{1}{2}\left[\Sigma_{V+A}(m_L,m_L)+\Sigma_{V-A}(m_L,m_L)\right]=\frac{1}{2}\tilde{\Sigma} \,,
\end{align}
The loop contribution from $N_{a/b}$ can also be computed as Eq.~\eqref{sigmamassandcoupling} with
\begin{align}
\frac{1}{2}\times&\left[\left((2g_L)_{ab}\right)^2+\left((2g_R)_{ab}\right)^2\right]\Sigma_{V+A}(M_a,M_b)={2}\left(|(g_L)_{ab}|^2+|(g_R)_{ab}|^2\right)\Sigma_{V+A}(M_a,M_b),\notag\\
\frac{1}{2}\times&2\;\text{Re}\left((2g_L)_{ab}(2g_R)_{ab}^*\right)\Sigma_{V-A}(M_a,M_b)=4\text{Re}((g_L)_{ab}(g_R)_{ab}^*)\Sigma_{V-A}(M_a,M_b).
\end{align}

We can use the same method to compute the contribution from charged current, with the coupling replaced by $\frac{V_{1a}}{\sqrt{2}}P_L$, $\frac{V_{2a}^*}{\sqrt{2}}P_R$, and the mass variables $m_L$, $M_a$, then the result will be
\begin{equation}
\begin{split}
\left[\left(\frac{V_{1a}}{\sqrt{2}}\right)^2+\left(\frac{V_{2a}^*}{\sqrt{2}}\right)^2\right]\Sigma_{V+A}(m,M_a)=\frac{V_{1a}^2+V_{2a}^2}{{2}}\Sigma_{V+A}(m,M_a),\\
2\;\text{Re}\left[\frac{V_{1a}}{\sqrt{2}}\cdot\left(\frac{V_{2a}^*}{\sqrt{2}}\right)^*\right]\Sigma_{V-A}(m,M_a)=\text{Re}(V_{1a}V_{2a})\Sigma_{V-A}(m,M_a).
\end{split}
\end{equation}



\newpage

\appendix

\noindent {\bf{Acknowledgements}}

This work is supported by the National Science Foundation of China (11775086).

\end{document}